\RequirePackage{fix-cm}
\documentclass[smallextended]{svjour3}       
\smartqed  
\usepackage{graphicx}
\begin{document}

\title{Constituent Counting Rules and Exotic Hadrons\thanks{Supported
    in part by the U.S.\ National Science Foundation under Grant No.\
    PHY-1403891.}  }



\author{Richard F. Lebed}


\institute{R.F. Lebed \at
              Department of Physics \\
              Arizona State University \\
              Tempe, AZ, USA 85287-1504 \\
              \email{richard.lebed@asu.edu}           
}

\date{January, 2018. \\ \\
The final authenticated version, published in
{\it Few-Body Systems}, is available online at
https://doi.org/10.1007/s00601-018-1427-2.}

\maketitle

\begin{abstract}
The constituent counting rules, {\it i.e.}, the scaling behavior of
amplitudes (in terms of the number of fundamental constituents) for
exclusive processes when high energy scales are present, have been
known for decades, and have been borne out in a number of experiments.
Such scaling would be sensitive, in particular, to possible exotic
multiquark content.  Here we examine how one may use the rules to test
for pentaquarks in electroproduction, or for tetraquarks in $e^+ e^-$
annihilation.  An interesting new type of scaling (separate Mandelstam
$s$ and $t$ behavior) arises in the forward scattering direction.  The
correct scaling arises naturally in AdS/QCD, in which the amplitudes
can be computed explicitly.
\keywords{QCD scaling rules \and Exotic hadrons}
\end{abstract}

\section{Introduction: Multiquark Hadrons}
\label{sec:Intro}

After more than 50 years of the quark model and more than 40 years of
QCD, we find ourselves at last in the era of multiquark hadrons.  The
first such unambiguous candidate, the $X(3872)$, was discovered by
BELLE in 2003~\cite{Choi:2003ue}, and has been established as a
primarily $c\bar c q\bar q$ state ($q$ a light quark).  Its existence
(with a well-defined mass and width) has been verified at multiple
experiments (BaBar, CDF, D\O, BESIII, LHCb, CMS, COMPASS), but the
unfortunate decline~\cite{Schumacher:2005wu} of the ostensible
pentaquark state $\Theta^+(1540)$ seen around the same time led the
physics community as a whole to be less confident in the acceptance of
$X(3872)$ as a true exotic state.  Numerous charmoniumlike (and
bottomoniumlike) states---some neutral and some charged---were
discovered in subsequent years.  However, not until 2014 with the LHCb
discovery of the charged $Z_c^-(4430)$ $c\bar c d\bar u$
state~\cite{Aaij:2014jqa}, in which was observed the looping Argand
diagram production amplitude behavior characteristic of a resonant
state, were most of the remaining multiquark-state skeptics persuaded.

As of the time of this writing, 35 heavy-quark exotics have been
observed: 30 in the charmoniumlike sector, 4 in the bottomoniumlike
sector, and 1 [$X(5568)$] believed to be $\bar b s d\bar
u$~\cite{D0:2016mwd}.  Most are tetraquarks, but 2 discovered by
LHCb~\cite{Aaij:2015tga} (called $P_c^+$) have pentaquark quantum
numbers.  The discovery history and status of these heavy-quark
exotics is summarized in the recent review~\cite{Lebed:2016hpi}, and
various proposals for future research directions are discussed
in~\cite{Briceno:2015rlt}.  Several other reviews have recently
appeared~\cite{Chen:2016qju,Esposito:2016noz,Guo:2017jvc,Ali:2017jda,Olsen:2017bmm,Karliner:2017qhf}.

It is especially interesting that, after decades of searches, the
first clear signals for multiquark hadrons appeared in the heavy-quark
sector.  In states of this sector, quantum-mechanical potential models
characterize -onium states very well and heavy-light states reasonably
well, exposing exotic states as ones that are superfluous in the
spectrum or have anomalous properties.  One may then hope that signals
of exotic states might appear in the $s$-quark sector, such as the
possibility of $\gamma p \! \to \! \phi
p$~\cite{Dey:2014tfa,Dey:2014npa} revealing hints of $s\bar s uud$
pentaquarks~\cite{Lebed:2015fpa}.

\section{Constituent Counting Rules}
\label{sec:Counting}

This paper focuses upon the use of one of the earliest diagnostics of
QCD, the constituent counting
rules~\cite{Brodsky:1973kr,Matveev:1973ra} developed in its first
decade, and specifically upon their use to reveal the multiquark
nature of heavy-quark states, as recently discussed in
Refs.~\cite{Brodsky:2015wza,Brodsky:2016uln,Brodsky:2017icd}.  The
counting rules are an expression of the approximate conformality and
scale independence of QCD at high energies; formally, they give the
leading-order scaling behavior of scattering amplitudes in terms of
the twist dimension of the operator responsible for the process
amplitude.  Obtaining the relevant scaling power is straightforward
when one uses the leading Feynman diagram for the process: Namely,
every propagator and vertex through which flows a large momentum
transfer [so that the overall process involves scattering through a
finite center-of-momentum (c.m.) frame angle $\theta_{\rm CM}$, {\it
i.e.}, fixed $t/s$] contributes to the leading-twist operator.
Supposing that this large scale is uniformly Mandelstam $s$---the
total c.m.\ energy for the process---then the invariant amplitude
${\cal M}$ and differential cross section $d\sigma/dt$ scale very
simply in terms of the total number (initial plus final) of
fundamental constituents (quarks, leptons, gauge bosons) $n$:
\begin{equation} \label{eq:count1}
{\cal M} \sim s^{-\frac n 2 + 2} \, , \ \ \ \ \frac{d\sigma}{dt} \sim
\frac{1}{s^{n-2}} \, .
\end{equation}
These scaling rules are genuinely rigorous and survive, up to $\ln s$
corrections, such varied effects as $\alpha_s$ running and
renormalization-group effects, Sudakov logarithm resummations, pinch
singularities, and endpoint singularities (see~\cite{Brodsky:2015wza}
for relevant references).

Moreover, numerous processes studied in experiments dating back
decades confirm the scaling predicted by Eq.~(\ref{eq:count1}),
$d\sigma \! /dt \! \sim \! s^{-8}$ for meson-baryon scattering and
$d\sigma \! /dt \! \sim s^{-10}$ for baryon-baryon
scattering~\cite{White:1994tj}, provided one considers ``large $s$''
to begin between some 100's of MeV to 2~GeV above threshold ({\it
i.e.}, beyond the resonant region), depending upon the specific
process.  As one particular example, Fig.~8 of
Ref.~\cite{Kawamura:2013iia} displays rather convincing experimental
evidence that $s^8 d\sigma \! /dt$ for $\pi^- p \! \to \! K^0 \Lambda$
approaches a constant beyond $s^{1/2}
\! \approx \! 2$~GeV\@.

\section{Constituent Counting Rules and Exotics}
\label{sec:Exotics}

The main thrust of Ref.~\cite{Kawamura:2013iia}, however, was to test
the oft-touted possibility ({\it e.g.}, in \cite{Guo:2017jvc}) that
$\Lambda(1405)$ is a (5-quark) $KN$ molecule because it is so light
even when compared to the nonstrange $N^*$'s such as $N(1535)$.
Applying Eq.~(\ref{eq:count1}) to such a state yields
\begin{eqnarray}
\frac{d\sigma}{dt} \left[ \pi^- p \to K^0 \Lambda (1405) \right] &
\sim & s^{-10} \, , \\
\frac{d\sigma}{dt} \left[ \gamma \, p \to K^0 \Lambda (1405) \right]
& \sim & s^{-9} \, .
\end{eqnarray}
The latter process, especially, is custom-made for measurement at
Jefferson Lab.

If such unique signatures are possible for light exotic candidate
states such as $\Lambda(1405)$, then, as Ref.~\cite{Brodsky:2015wza}
reasoned, similar results should hold for the heavy-quark exotics as
well.  In particular, if $Z_c^\pm$ is a tetraquark state, then
\begin{eqnarray}
\frac{d\sigma}{dt} \left[ e^+ e^- \to Z_c^\pm \pi^\mp \right] & \sim &
s^{-6} \, , \nonumber \\
\frac{d\sigma}{dt} \left[ e^+ e^- \to Z_c^+ Z_c^- \right] & \sim &
s^{-8} \, . \label{eq:dsigmadtZc}
\end{eqnarray}
On the other hand, if the quarks in $Z_c^+(c\bar c u \bar d)$ (or its
charge conjugate $Z_c^-$) are arranged into two diquarks $(cu)$ and
$(\bar c \bar d)$ that are tightly bound and therefore scatter as
intact units, then
\begin{eqnarray}
\frac{d\sigma}{dt} \left[ e^+ e^- \to Z_c^\pm \pi^\mp \right] & \sim &
s^{-4} \, , \nonumber \\
\frac{d\sigma}{dt} \left[ e^+ e^- \to Z_c^+ Z_c^- \right] & \sim &
s^{-4} \, . \label{eq:dsigmadtZcdiq}
\end{eqnarray}
These scalings provide rather distinctive experimental signals.
However, one may object that they only apply at energies $\sim \!
\sqrt{s}$ rather high above threshold.  In order to remove systematic
differences due to the presence of nontrivial resonant regions, one
may consider ratios, {\it e.g.},
\begin{equation}
\frac{\sigma \left( e^+ e^- \to Z_c^\pm \pi^\mp \right)}
{\sigma \left( e^+ e^- \to \mu^+ \mu^- \right)} = \left| F_{Z_c}(s)
\right|^2 \propto \frac{1}{s^4} \, ,
\end{equation}
or $\propto \! s^{-2}$ if $Z_c$ is formed of tightly bound diquarks.
Here, $F_{Z_c}$ is the electromagnetic form factor of the $Z_c^\pm$.

One may also consider ratios such as
\begin{equation}
\frac{\sigma \left( e^+ e^- \to Z_c^+ \pi^- \right)}
{\sigma \left( e^+ e^- \to \Lambda_c^+ \bar \Lambda_c^- \right)} \, ,
\end{equation}
in which both final states contain the same number (6) of fundamental
constituents and the same heavy quarks ($c\bar c$), their effects
cancelling in the ratio.  In this case, the value of the ratio is a
direct measure of how the components of the $Z_c$ are assembled; one
may expect a smaller [but still $O(1)$] value if the $Z_c$ consists of
two loosely bound mesons.  The same types of counting considerations
apply to the electroproduction of hidden-charm pentaquarks, $e^- p \to
e^- P_c^+ (c \bar c uud)$, a process of great interest to JLab.

Another interesting exception to the scalings given above was noted in
Ref.~\cite{Guo:2016fqg}: The naive scalings require every constituent
of the process to participate in the scattering through a high
momentum transfer.  If some of them are produced softly with respect
to any of the energetic constituents---for example, in $Z_c$
production, the $c\bar c$ pair might be produced from the vacuum
through gluons emitted from one of the light quarks $q$ that are soft
in the $q$ rest frame---then the $1/s$ scaling occurs at a lower power
than advertised.  In this particular case, the $c\bar c$ production
amplitude is suppressed by $1/m_c$ due to the charm-quark propagator
in $gg$ fusion, while $c$ and $\bar c$ are created with virtuality
$m_c^2$ and therefore must exchange gluons of $q^2 \! \sim
\!  m_c^2$ in order to be promoted to becoming nearly on-shell $Z_c$
components.  One can check that this mechanism is not competitive with
ones in which the $c\bar c$ pair is created from the original hard
subprocess until $\sqrt{s} \! = \! O(20~{\rm GeV})$~\cite{BLL:2017}.

\section{New Constituent Counting Rules in the Forward Region}
\label{sec:Forward}

Interesting novel constituent counting rules arise when any of the
constituents undergoing hard momentum transfer are constrained to lie
along the forward direction of scattering~\cite{Brodsky:2016uln}.  As
noted above, traditional constituent counting rules require scattering
through a finite angle $\theta_{\rm CM}$.  Neglecting masses, one has
\begin{equation}
t, u = -\frac{s}{2} \left( 1 \mp \cos \theta_{\rm CM} \right) \, ,
\end{equation}
so that $s, t, u$ are all of the same order for finite $\theta_{\rm
  CM}$.  However, in the extreme forward (backward) direction, for
which $\theta_{\rm CM} \! = \! \varepsilon$ ($\theta_{\rm CM} \! = \!
\pi \! - \!  \varepsilon$) with $\varepsilon \! \ll \! 1$, then:
\begin{equation}
-t = s \frac{\varepsilon^2}{4} \ll s \approx -u \ \
\left( -u = s \frac{\varepsilon^2}{4} \ll s \approx -t \right) \, .
\end{equation}
The existence of a new scale (specifically, the hierarchy here reads
$\Lambda_{\rm QCD}^2 \! \ll \! -t \! \ll \! s$) implies the existence
of new scaling laws for cross sections.

\begin{figure}[ht]
\begin{center}
  \includegraphics[width=0.75\textwidth]{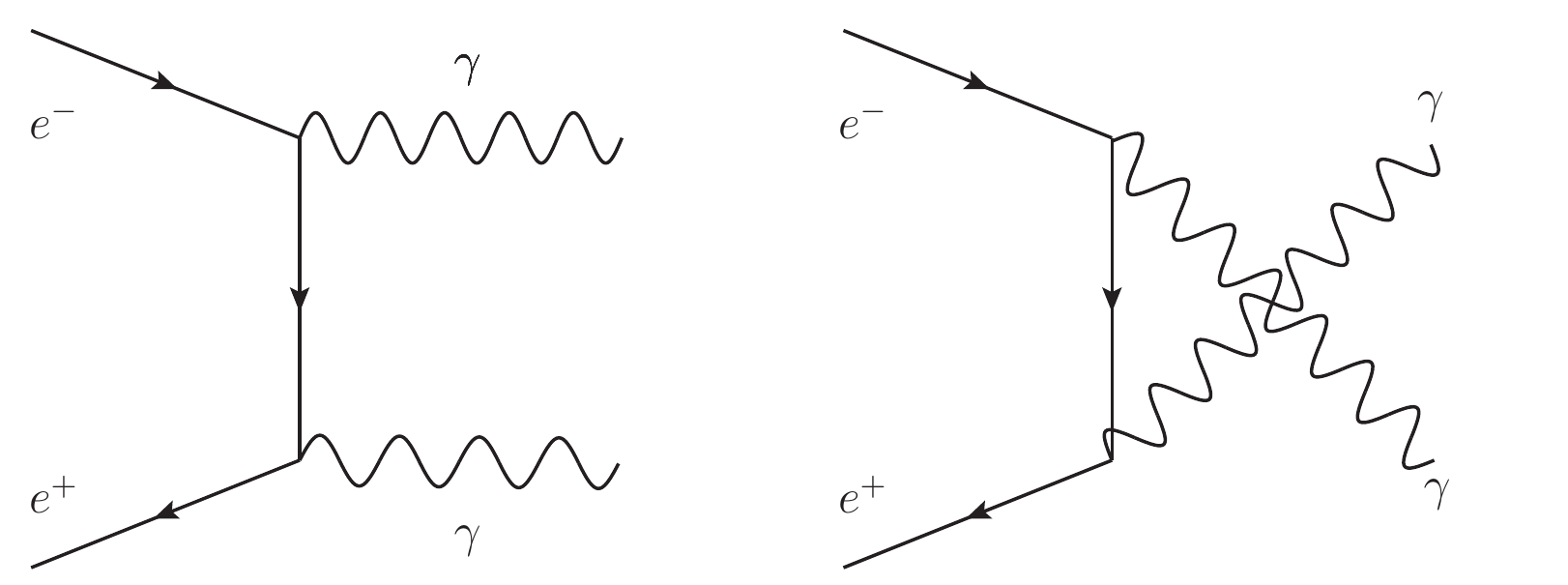}
\end{center}
\caption{Leading-order diagrams ($t$- and $u$-channel, respectively)
  for $e^+ e^- \! \to \! \gamma \gamma$.}
\label{fig:eegamgam}
\end{figure}

Consider the textbook example~\cite{Peskin:1995ev} of $e^+ e^- \! \to
\! \gamma^* \! \to \! q\bar q \! \to \! {\rm hadrons}$, whose usual
scale-invariant cross section goes as $\alpha_{\rm EM}^2 / s$.  The
$1/s$ factor arises through a combination of the photon propagator,
fermion traces, and phase space.  But one can also show for the
process $e^+ e^- \! \to \! \gamma \gamma$, which has $t$ (forward) and
$u$ (backward) singularities due to the lepton propagator (see
Fig.~\ref{fig:eegamgam}), that in fact its cross section {\em also\/}
scales as $1/s$~\cite{Davier:2006fu}---specifically, as $\alpha_{\rm
EM}^2/s|t|$ in the forward direction.

It was further argued in Ref.~\cite{Brodsky:2016uln} that additional
suppressions by $\alpha_{\rm EM}^{\vphantom{2}} \kappa^2/|t|$ or
$\alpha_{\rm EM}^2 \kappa^4/|t|^2$, respectively [where $\kappa \! =
\! O(\Lambda_{\rm QCD})$], arise when one or both of the photons
convert to neutral vector mesons $V^0$, are the cost of constraining
the $q\bar q$ components of $V^0$ both to propagate in the forward
direction.  However, this result must be reassessed in light of the
objection noted above, that obtaining the full naive scaling behavior
requires {\em all\/} relevant lines in the diagram to carry large
momentum transfers; if not, then the twist of the corresponding
operator is smaller~\cite{Brodsky:2017icd}.  Such is the case
(anticipated in~\cite{Guo:2016fqg}) when the photon couples solely to
a single $V^0$ of fixed squared mass $m_{V^0}^2 \! \equiv \! s_1$;
then the photon---which generically would have $O(|t|)$
virtuality---carries a precisely fixed virtuality, $s_1$.  Since the
photon propagator is one of the contributors to the total counting
rule, the scaling law is modified: The $\kappa^2/|t|$ suppressions are
replaced by $O(1)$ factors of $\kappa^2/s_1$.  In this exceptional
configuration, one recovers the expectation of {\it vector meson
dominance\/} (VMD): The $V^0$ behaves as a single fundamental
constituent.

Away from this special point, the original expectations
of~\cite{Brodsky:2016uln} remain valid~\cite{Brodsky:2017icd}.  In
particular, the inclusive forward cross section $\sigma(e^+ e^- \! \to
\! \gamma \gamma^* \! \to \! \gamma q \bar q)$ scales as $1/|t|^2$,
while $V^0$ created by hadronic processes ({\it e.g.}, in $pp$
scattering) no longer behave as though composed of a single
constituent.

\section{Explicit Checks in AdS/QCD}
\label{sec:AdSQCD}

One may observe the explicit manifestation of the constituent counting
rules in AdS/QCD, which is a model for QCD that treats strongly
coupled confining theories as dual to weakly coupled gravity theories
in 5 spacetime dimensions, in such a way that confinement is
represented by the familiar 4-dimensional fields only being able to
penetrate a finite distance $z$ into the 5$^{\rm th}$ {\it bulk\/}
dimension.  In the particular variant known as the {\it soft-wall
AdS/QCD light-front model\/}~\cite{Brodsky:2007hb,Branz:2010ub}, each
coupling is weighted by the Gaussian factor $e^{-\kappa^2 z^2}$, and
then all nonperturbative overlap integrals, such as form factors, can
be computed in closed form as confluent hypergeometric (Kummer)
functions or Bessel functions.

In particular, not only can one show that the square of the $\gamma^*
V^0$ transition form factor $|G_V(s_1)|^2$ scales as $\kappa^2/s_1$ as
discussed above, but in soft-wall AdS/QCD in the large-$s_1$ limit one
calculates exactly that:
\begin{equation}
\left| \frac{G_V(s_1)}{G_V(0)} \right|^2 = \frac{9\pi^3}{256}
\cdot \frac{\kappa^2}{s_1} \, .
\end{equation}

Returning to the issue of counting constituents, let us consider the
process $e^+ e^- \! \to \! H_{n_1} H_{n_2}$ where hadrons $H$ contain
$n_1$ and $n_2$ constituents, respectively.  Then the form factors $F$
in
\begin{equation}
\frac{\sigma \left( e^+ e^- \to H_{n_1} H_{n_2} \right)}
{\sigma \left( e^+ e^- \to \mu^+ \mu^- \right)} = \left| F_{H_{n_1}
H_{n_2}} (s) \right|^2 \, ,
\end{equation}
are computed in AdS/QCD as the overlap integrals:
\begin{equation}
F_{H_{n_1} H_{n_2}} (Q^2) = \int_0^\infty \! dz \, V(Q,z) \,
\phi_{n_1} (z) \, \phi_{n_2} (z) \, ,
\end{equation}
where $V(Q,z) \! = \! \Gamma (1+a) U(a,0,\kappa^2 z^2)$ is the {\it
vector-field bulk-to-boundary propagator}, $a \! \equiv \!
Q^2/4\kappa^2$, $U$ is a Kummer function, and the {\it bulk
profiles\/} of the hadrons are given by
\begin{equation}
\phi_n (z) = \sqrt{\frac{2}{\Gamma(n-1)}} \kappa^{n-1} z^{n-3/2}
e^{-\kappa^2 z^2 / 2} \, .
\end{equation}
One obtains the closed-form result
\begin{equation}
F_{H_{n_1} H_{n_2}} (Q^2) = \frac{\Gamma \left( \frac{n_1 + n_2}{2}
\right) \Gamma \left( \frac{n_1 + n_2}{2} - 1 \right)} {\sqrt{\Gamma
(n_1 - 1)} \sqrt{\Gamma (n_2 - 1)}} \frac{\Gamma(a+1)} {\Gamma \left( a
+ 1 + \frac{n_1 + n_2}{2} - 1 \right)} \, .
\end{equation}

Using the large-$a$ ($Q^2 \! \to \! s$) expansion of $\Gamma$
functions, one immediately has
\begin{eqnarray}
F_{Z_c^+ \pi^-} \sim s^{-2} & \Longrightarrow & \frac{d\sigma}{dt}
\left[ e^+ e^- \to Z_c^+ \pi^- \right] \sim s^{-6} \, , \nonumber \\
F_{Z_c^+ Z_c^-} \sim s^{-3} & \Longrightarrow & \frac{d\sigma}{dt}
\left[ e^+ e^- \to Z_c^+ Z_c^- \right] \sim s^{-8} \, ,
\end{eqnarray}
exactly as predicted by Eqs.~(\ref{eq:dsigmadtZc}).  If the components
of $Z_c$ are treated as fundamental diquarks, then
Eqs.~(\ref{eq:dsigmadtZcdiq}) immediately follow.

\section{Conclusions}
\label{sec:Concl}

Exotic hadrons, at least in the form of tetraquarks and pentaquarks,
are here to stay: Over 30 such states ($X, Y, Z, P_c$) have thus far
been observed.  Producing and characterizing them and their decay
modes will be one of the major thrusts of experimental facilities: not
just the existing LHCb, BESIII, and Belle(II) programs, but at JLab,
COMPASS, PANDA at FAIR, and elsewhere.

The old and well-known constituent counting rules provide simple and
straightforward tests of exoticity: Scalings of both cross sections at
high energies and ratios of cross sections at energies all the way
down to threshold provide important handles on exotic versus nonexotic
structure.  New forward-scattering ($s \! \gg \! |t| \! \gg \!
\Lambda_{\rm QCD}^2$) constituent counting rules provide yet more
useful experimental tests of hadronic substructure.  One must,
however, in all processes confirm the expected scaling behavior based
upon the leading twist dimension/number of propagators and vertices
through which flows large momentum transfers.

One may also use AdS/QCD, in which the nonperturbative amplitudes can
be computed explicitly, to obtain expressions in which the
expectations of the constituent counting rules are explicitly
manifested.

\begin{acknowledgements}
  I gratefully acknowledge Stan Brodsky and Valery Lyubovitskij for
  the successful collaboration that led to the publications described
  here.
\end{acknowledgements}


\begin{thebibliography}{}
%
\bibitem{Choi:2003ue} 
  S.-K.~Choi {\it et al.} [Belle Collaboration], {\em Observation of a
  Narrow Charmonium-Like State in Exclusive $B^\pm \to K^\pm \pi^+
  \pi^- \! J/\psi$ Decays},
  Phys.\ Rev.\ Lett.\ {\bf 91}, 262001 (2003)
  [hep-ex/0309032].
%
\bibitem{Schumacher:2005wu} 
  R.A.~Schumacher,
  {\em The Rise and Fall of Pentaquarks in Experiments},
  AIP Conf.\ Proc.\  {\bf 842}, 409 (2006)
  [nucl-ex/0512042].
%
\bibitem{Aaij:2014jqa} 
  R.~Aaij {\it et al.} [LHCb Collaboration],
  {\em Observation of the Resonant Character of the $Z(4430)^-$
  State},
  Phys.\ Rev.\ Lett.\  {\bf 112}, 222002 (2014)
  [arXiv:1404.1903 [hep-ex]].
%
\bibitem{D0:2016mwd} 
  V.M.~Abazov {\it et al.} [D0 Collaboration],
  {\em Evidence for a $B_s^0 \pi^\pm$ State},
  Phys.\ Rev.\ Lett.\  {\bf 117}, 022003 (2016)
  [arXiv:1602.07588 [hep-ex]].
%
\bibitem{Aaij:2015tga} 
  R.~Aaij {\it et al.} [LHCb Collaboration],
  {\em Observation of $J \! / \! \psi \, p$ Resonances Consistent with
  Pentaquark States in $\Lambda_b^0 \to J \! / \! \psi \, K^- p$
  Decays},
  Phys.\ Rev.\ Lett.\  {\bf 115}, 072001 (2015)
  [arXiv:1507.03414 [hep-ex]].
%
\bibitem{Lebed:2016hpi} 
  R.F.~Lebed, R.E.~Mitchell, and E.S.~Swanson,
  {\em Heavy-Quark QCD Exotica},
  Prog.\ Part.\ Nucl.\ Phys.\  {\bf 93}, 143 (2017)
  [arXiv:1610.04528 [hep-ph]].
%
\bibitem{Briceno:2015rlt} 
  R.A.~Brice\~no {\it et al.},
  {\em Issues and Opportunities in Exotic Hadrons},
  Chin.\ Phys.\ C {\bf 40}, 042001 (2016)
  [arXiv:1511.06779 [hep-ph]].
%
\bibitem{Chen:2016qju} 
  H.-X.~Chen, W.~Chen, X.~Liu, and S.-L.~Zhu,
  {\em The Hidden-Charm Pentaquark and Tetraquark States},
  Phys.\ Rept.\  {\bf 639}, 1 (2016)
  [arXiv:1601.02092 [hep-ph]].
%
\bibitem{Esposito:2016noz} 
  A.~Esposito, A.~Pilloni, and A.D.~Polosa,
  {\em Multiquark Resonances},
  Phys.\ Rept.\  {\bf 668}, 1 (2016)
  [arXiv:1611.07920 [hep-ph]].
%
\bibitem{Guo:2017jvc} 
  F.-K.~Guo, C.~Hanhart, U.-G.~Mei{\ss}ner, Q.~Wang, Q.~Zhao, and
  B.-S.~Zou,
  {\em Hadronic Molecules},
  Rev.\ Mod.\ Phys.\  {\bf 90}, 015004 (2018)
  [arXiv:1705.00141 [hep-ph]].
%
\bibitem{Ali:2017jda} 
  A.~Ali, J.S.~Lange, and S.~Stone,
  {\em Exotics: Heavy Pentaquarks and Tetraquarks},
  Prog.\ Part.\ Nucl.\ Phys.\  {\bf 97}, 123 (2017)
  [arXiv:1706.00610 [hep-ph]].
%
\bibitem{Olsen:2017bmm} 
  S.L.~Olsen, T.~Skwarnicki, and D.~Zieminska,
  {\em Non-Standard Heavy Mesons and Baryons, an Experimental Review},
  Rev.\ Mod.\ Phys.\  {\bf 90}, 015003 (2018)
  [arXiv:1708.04012 [hep-ph]].
%
\bibitem{Karliner:2017qhf} 
  M.~Karliner, J.L.~Rosner, and T.~Skwarnicki,
  {\em Multiquark States},
  arXiv:1711.10626 [hep-ph].
%
\bibitem{Dey:2014tfa} 
  B.~Dey {\it et al.} [CLAS Collaboration],
  {\em Data Analysis Techniques, Differential Cross Sections, and Spin
  Density Matrix Elements for the Reaction $\gamma p \rightarrow \phi
  p$},
  Phys.\ Rev.\ C {\bf 89}, 055208 (2014);
  addendum: [Phys.\ Rev.\ C {\bf 90}, 019901 (2014)]
  [arXiv:1403.2110 [nucl-ex]].
%
\bibitem{Dey:2014npa} 
  B.~Dey,
  {\em Phenomenology of $\phi$ Photoproduction from Recent CLAS Data
  at Jefferson Lab},
  arXiv:1403.3730 [hep-ex].
%
\bibitem{Lebed:2015fpa} 
  R.F.~Lebed,
  {\em Diquark Substructure in $\phi$ Photoproduction},
  Phys.\ Rev.\ D {\bf 92}, 114006 (2015)
  [arXiv:1510.01412 [hep-ph]].
%
\bibitem{Brodsky:1973kr} 
  S.J.~Brodsky and G.R.~Farrar,
  {\em Scaling Laws at Large Transverse Momentum},
  Phys.\ Rev.\ Lett.\  {\bf 31}, 1153 (1973).
%
\bibitem{Matveev:1973ra} 
  V.A.~Matveev, R.M.~Muradian, and A.N.~Tavkhelidze,
  {\em Automodellism in the Large-Angle Elastic Scattering and
  Structure of Hadrons},
  Lett.\ Nuovo Cim.\  {\bf 7}, 719 (1973).
%
\bibitem{Brodsky:2015wza} 
  S.J.~Brodsky and R.F.~Lebed, {\em QCD Dynamics of Tetraquark
  Production}, Phys.\ Rev.\ D {\bf 91}, 114025 (2015)
  [arXiv:1505.00803 [hep-ph]].
%
\bibitem{Brodsky:2016uln} 
  S.J.~Brodsky, R.F.~Lebed, and V.E.~Lyubovitskij,
  {\em QCD Compositeness As Revealed in Exclusive Vector Boson
  Reactions through Double-Photon Annihilation: $e^+ e^- \to \gamma
  \gamma^\ast \to \gamma V^0 $ and $e^+ e^- \to \gamma^\ast
  \gamma^\ast \to V^0 V^0$},
  Phys.\ Lett.\ B {\bf 764}, 174 (2017)
  [arXiv:1609.06635 [hep-ph]].
%
\bibitem{Brodsky:2017icd}
  S.J.~Brodsky, R.F.~Lebed, and V.E.~Lyubovitskij,
  {\em QCD Constituent Counting Rules for Neutral Vector Mesons},
  Phys.\ Rev.\ D {\bf 97}, 034009 (2018)
  [arXiv:1712.08853 [hep-ph]].
%
\bibitem{White:1994tj} 
  C.~White {\it et al.},
  {\em Comparison of 20 Exclusive Reactions at Large $t$},
  Phys.\ Rev.\ D {\bf 49}, 58 (1994).
%
\bibitem{Kawamura:2013iia} 
  H.~Kawamura, S.~Kumano, and T.~Sekihara,
  {\em Determination of Exotic Hadron Structure by
  Constituent-Counting Rule for Hard Exclusive Processes},
  Phys.\ Rev.\ D {\bf 88}, 034010 (2013)
  [arXiv:1307.0362 [hep-ph]].
%
\bibitem{Guo:2016fqg} 
  F.-K.~Guo, U.-G.~Mei{\ss}ner, and W.~Wang,
  {\em On the Constituent Counting Rule for Hard Exclusive Processes
  Involving Multi-Quark States},
  Chin.\ Phys.\ C {\bf 41}, 053108 (2017)
  [arXiv:1607.04020 [hep-ph]].
%
\bibitem{BLL:2017}
  S.J.~Brodsky and V.E.~Lyubovitskij, private communication.
%
\bibitem{Peskin:1995ev} 
  M.E.~Peskin and D.V.~Schroeder,
  {\em An Introduction to Quantum Field Theory},
  Westview Press, New York (1995).
%
\bibitem{Davier:2006fu} 
  M.~Davier, M.E.~Peskin, and A.~Snyder,
  {\em Two-Photon Exchange Model for Production of Neutral Meson Pairs
  in $e^+ e^-$ Annihilation},
  hep-ph/0606155.
%
\bibitem{Brodsky:2007hb} 
  S.J.~Brodsky and G.F.~de~Teramond,
  {\em Light-Front Dynamics and AdS/QCD Correspondence: The Pion Form
  Factor in the Space- and Time-Like Regions},
  Phys.\ Rev.\ D {\bf 77}, 056007 (2008)
  [arXiv:0707.3859 [hep-ph]].
%
\bibitem{Branz:2010ub} 
  T.~Branz, T.~Gutsche, V.E.~Lyubovitskij, I.~Schmidt, and A.~Vega,
  {\em Light and Heavy Mesons in a Soft-Wall Holographic Approach},
  Phys.\ Rev.\ D {\bf 82}, 074022 (2010)
  [arXiv:1008.0268 [hep-ph]].
%
\end{thebibliography}
\end{document}